# Theory of Carry Value Transformation (CVT) and it's Application in Fractal formation

**Pal Choudhury Pabitra[1], Sahoo Sudhakar [1], Nayak Birendra Kumar[2], and Hassan Sk. Sarif [2]**

[1,] Applied Statistics Unit, Indian Statistical Institute, Kolkata, 700108, INDIA
[2.] P.G. Department of Mathematics, Utkal University, Bhubaneswar-751004

Email: pabitrapalchoudhury@gmail.com, sudhakar.sahoo@gmail.com, bknatuu@yahoo.co.uk & sarimif@gmail.com

**Abstract**. In this paper the theory of Carry Value Transformation (CVT) is designed and developed on a pair of n-bit strings and is used to produce many interesting patterns. One of them is found to be a self-similar fractal whose dimension is same as the dimension of the Sierpinski triangle. Different construction procedures like L-system, Cellular Automata rule, Tilling for this fractal are obtained which signifies that like other tools CVT can also be used for the formation of self-similar fractals. It is shown that CVT can be used for the production of periodic as well as chaotic patterns. Also, the analytical and algebraic properties of CVT are discussed. The definition of CVT in two-dimension is slightly modified and its mathematical properties are highlighted. Finally, the extension of CVT and modified CVT (MCVT) are done in higher dimensions.

## 1. Introduction

Benoit Mandelbrot coined the word fractal from the Latin adjective fractus. The corresponding Latin verb frangere means 'to break' to create irregular fragments. The precise definition of "Fractal" according to Benoit Mandelbrot is as a set for which the Hausdroff Besicovitch dimension strictly exceeds the topological dimension [1].

Many things in nature are very complex, chaotic but exhibit some self-similarity. The complexity of fractals and the property of self-similarity have a large set of real world applications. Fractals can be generated using construction procedure/algorithms/simply by repetition of mathematical formula that are often recursive and ideally suited to computer.

In this paper a new transformation named as Carry Value Transformation (CVT) is defined in binary number system and using CVT we are trying to explore that behind the complexity of nature there remains a simple methodology about which most often we are ignorant. Earlier, in [2] we have used CVT with Cellular Automata in efficient hardware design of some basic arithmetic operations. But in this paper although we emphasize the formation of self-similar-fractals, the algorithm using CVT also produces periodic and chaotic patterns.

The underlying development of CVT is same as the concept of Carry Save Adder (CSA) [3] where carry or overflow bits generated in the addition process of two integers are saved in the memory. Here we perform the bit wise XOR operation of the operands to get a string of sum-bits (ignoring the carry-in) and simultaneously the bit wise ANDing of the operands to get a string of carry-bits, the latter string is padded with a '0' on the right to signify that there is no carry-in to the LSB.

The organization of the paper is as follows. Section II discusses some of the basic concepts on fractals, L-systems, tilling problem, Cellular Automata etc. which are used in the subsequent sections. The concept of CVT is defined in section III. It can be found in section IV that CVT generates a beautiful self-similar fractal whose dimension is found to be same as that of monster fractal, Sierpinski triangle. Section V deals with the various ways like L-System, Cellular Automata and Tillings by which the same fractal in binary number system can be constructed. CVT can also be used for the production of periodic as well as chaotic patterns are shown in section VI. The analytical and algebraic properties of CVT are discussed in section VII. The definition of CVT in two-dimension is slightly modified and its mathematical properties are highlighted in section VIII. Section IX deals with the extension of CVT and modified CVT (MCVT) in higher dimensions. On highlighting other possible applications of CVT and some future research directions a conclusion is drawn in section X.

It should be noted that a preliminary version of this paper has been published in an international conference [13].

## 2. Fractal Basics

This sections deals with the basics of Fractals [1, 4, 5] and the various ways the fractals may be constructed like L-Systems [6], Cellular Automata [7, 8] and synthesis of Tilling [9] etc. It is intended for readers who are not conversant with the fundamentals of these concepts.

2.1 *Why we study fractals*
We feel very much worried due to our inability to describe using the traditional Euclidean Geometry, the shape of cloud, a mountain, a coastline or a tree. In nature, clouds are not really spherical, mountains are not conical, coastlines are not circular, even the lightning doesn't travel in a straight line. More generally, we could be able to conclude that many patterns of nature are so irregular and fragmented, that, compared with *Euclid Geometry* –a term, can be used in this regard to denote all of standard geometry. Mathematicians have over the years disdained this challenge and have increasingly chosen to flee nature by devising theories unrelated to natural objects we can see or feel.
After a long time, responding to this challenge, Benoit Mandelbrot developed a new geometry of nature and implemented its use in a number of diverse arenas of science such as Astronomy, Biology, Mathematics, Physics, and Geography and so on [1, 4, 5, 10, 11]. This new-born geometry can describe many of the irregular and fragmented (chaotic) patterns around us, and leads to full-fledged theories, by identifying a family of shapes, now-a-days which we people call 'FRACTALS'.

2.2 *Measuring Fractal dimension*
The fractal dimension alone does not give an idea of what "fractals" are really about Mandelbrot founded his insights in the idea of self similarity, requiring that a true fractal "fracture" or break apart into smaller pieces that resemble the whole. This is a special case of the idea that there should be a dynamical system underlying the geometry of the set. This is partly why the idea fractals have become so popular throughout science; it is a fundamental aim of science to seek to understand the underlying dynamical properties of any natural phenomena. It has now become apparent that relatively simple dynamics, more precisely dynamical system can produce the fantastically intricate shapes and behavior that occur throughout nature.
Now let us try to define what fractal dimension (self-similarity dimension) is. Given a self-similar structure [1], there is a relation between the reduction factor (scaling factor) 'S' and the number of pieces 'N' into which the structure can be divided; and that relation is…
$N = 1/S^D$, equivalently, $D = \log(N)/\log(1/S)$
This 'D' is called the Fractal dimension (Self-similarity dimension)

2.3 *Ways to construct fractals*

  2.3.1 *Lindenmayer Systems (L-system) produces fractals*
As a biologist, Aristid Lindenmayer [6] studied growth patterns in various types of algae. In 1968 he developed Lindenmayer systems (or L-Systems) as a mathematical formalism for describing the growth of simple multi-cellular organisms. The central concept of L-System is that of rewriting. In general rewriting is a technique for defining complex object by successively replacing parts of a simple initial object using a set of production rules.
Definition of an L-System:
An L-system is a formal grammar consisting of 4 parts:
A set of *variables*: symbols that can be replaced by production rules.
A set of *constants*: symbols that do not get replaced.
An *axiom*, which is a string, composed of some number of variables and/or constants. The axiom is the initial state of the system.
A set of *production rules* defining the way variables can be replaced with combinations of constants and other variables. A production consists of two strings - the predecessor and the successor.

2.3.1 *Fractals by Cellular Automata rules*
The scientific output of Wolfram's [7, 8] work played a central role in launching Cellular Automata (CA) as a new field of science to understand the complexity of nature. Starting from an initial seed he studied the space-time diagram of all the 256 three-neighborhood elementary CA rules and classified the rules into four distinct classes according to the complexity of the pattern. According to him the Class 2 rules deals with the periodic and fractal patterns.

In 1-D the *global state* or simply *state* of a CA at any time-instant $t$ is represented as a vector $X^t = (x_1^t, x_2^t, \ldots, x_n^t)$ where $x_i^t$ denotes the bit in the $i^{th}$ cell $x_i$ at time-instant $t$. If the "present state" of an $n$-bit CA (at time $t$) is $X^t$, its "next state" (at time $t+1$), denoted by $X^{t+1}$, is in general given by the *global mapping* $F(X^t) = (f^1(lb^t, x_1^t, x_2^t), f^2(x_1^t, x_2^t, x_3^t), \ldots, f^n(x_{n-1}^t, x_n^t, rb^t))$, where $f^i$ is a local mapping to the $i^{th}$ cell and $lb$ and $rb$ denote respectively the left boundary of $x_1$ and right boundary of $x_n$ incase of periodic boundary CA and those values are 0 in case of null boundary CA. If the same local mapping (rule) determines the "ne*x*t" bit in each cell of a CA, the CA will be called a Uniform CA, otherwise it will be called a Hybrid CA.

For our purpose, we have used one-way *CA*, which allows only one-way communication, i.e., in a 1-D array each cell depends only on itself and its left neighbor. One can also consider dependence on the cell and its right neighbor. However both sides dependence is not allowed.

Just like L-system and Cellular Automata, fractals can also be obtained by Iterated Function Systems [4] and using different Tiles [9]. A complex figure can be easily (in most of the cases) synthesized by using of tiles. Next section discusses a new and efficient construction tool named as CVT by which uncountable number of fractal patterns can be generated.

## 3. Carry Value Transformation (CVT)

The carry or overflow bits are usually generated at the time of addition between two n-bit strings. In the usual addition process, carry value is always a single bit and if generated then it is added column wise with other bits and not saved in its own place. But the carry value defined here are the usual carries generated bit wise and stored in their respective places as shown in "Fig. 1".

$$\text{carry value} = c_n \quad c_{n-1} \ldots\ldots\ldots\ldots\ldots\ldots c_1 \quad 0$$
$$a = a_n \quad a_{n-1} \ldots\ldots\ldots\ldots\ldots a_1$$
$$b = b_n \quad b_{n-1} \ldots\ldots\ldots\ldots\ldots b_1$$
$$a \oplus b = a_n \oplus b_n \quad a_{n-1} \oplus b_{n-1} \ldots\ldots\ldots a_1 \oplus b_1$$

[Figure 1: Carry generated in i$^{th}$ column is saved in (i-1)$^{th}$ column]

Thus to find out the carry value we perform the bit wise XOR operation of the operands to get a string of sum-bits (ignoring the carry-in) and simultaneously the bit wise ANDing of the operands to get a string of carry-bits, the latter string is padded with a '0' on the right to signify that there is no carry-in to the LSB. Thus the corresponding decimal value of the string of carry bits is always an even integer.

Now we can give a precise definition of CVT as follows:

Let $B = \{0,1\}$ and CVT is a mapping defined as $CVT : (B_n \times B_n) \to B_{n+1}$ where $B_n$ is the set of strings of length $n$ on $B = \{0,1\}$. More specifically, if $a = (a_n, a_{n-1}, \ldots, a_1)$ and $b = (b_n, b_{n-1}, \ldots, b_1)$ then $CVT(a,b) = (a_n \wedge b_n, a_{n-1} \wedge b_{n-1}, \ldots, a_1 \wedge b_1, 0)$ is an (n+1) bit string, belonging to set of non-negative integers, and can be computed bit wise by logical AND operation followed by a 0, which denotes no carry is generated in the LSB at the time of addition procedure. In other words, CVT is a mapping from where $\mathbb{Z}$ is set of non-negative integers.

*Illustration:*
Suppose, we want the CVT of the numbers $(13)_{10} \equiv (1101)_2$ and $(14)_{10} \equiv (1110)_2$. Both are 4-bit numbers. The carry value is computed as follows:

<u>Carry:   1 1 0 0 0</u>
<u>Augend:    1 1 0 1</u>
<u>Addend:    1 1 1 0</u>
XOR:       0 0 1 1

[Figure 2: Carry generated in i$^{th}$ column is saved in (i-1)$^{th}$ column]

Conceptually, in the general addition process the carry or overflow bit from each stage (if any) goes to the next stage so that, in each stage after the first (i.e. the LSB position with no carry-in), actually a 3-bit addition is performed instead of a 2-bit addition by means of the full adder. Instead of going for this traditional method, what we do is that we perform the bit wise XOR operation of the operands (ignoring the carry-in of each stage from the previous stage) and simultaneously the bit wise ANDing of the operands to get a string of carry-bits, the latter string is padded with a '0' on the right to signify that there is no carry-in to the LSB (the overflow bit of this ANDing being always '0' is simply ignored). In our example, bit wise XOR gives $(0011)_2 \equiv (3)_{10}$ and bit wise ANDing followed by zero-padding gives $(11000)_2 \equiv (24)_{10}$. Thus $CVT(1101,1110) = 11000$ and equivalently in decimal notation one can write $CVT(13,14) = 24$. In the next section we have used the carry value in decimal to construct the CV table.

**4. Generation of Self-fractal using CVT**

A table is constructed that contains only the carry values (or even terms) defined above between all possible integers a's and b's arranged in an ascending order of x and y-axis respectively. We observe some interesting patterns in the table. We would like to make it clear how the CV-table is constructed.

**Step 1**. Arrange all the integers 0 1 2 3 4 5 6 ... (as long as we want) in ascending order and place it in both, uppermost row and leftmost column in a table.

**Step 2**. Compute $CVT(a,b)$ as mentioned in section III and store it in decimal form in the (a, b) position.

**Step 3**. Then we look on the pattern of any integer, and we have made it color. This shows a very beautiful consistent picture, which we see as a fractal as shown in table I followed by "Fig. 3".

Choosing different sequence of rows and columns in the CV- table in step 1, one can obtain uncountable number of patterns only by computing a single operation $CVT(a,b)$ in each entry position (a, b) of the table. Thus one of the advantages of CVT over other construction tools lies in the number of cell evolutions for the formation of fractals. In this case only one uniform evolution is required at each cell position to obtain the required fractal pattern where as in case of L-System, Cellular Automata, Iterated Function System etc many iterations are required.

*Illustration:*

[Figure 3: A fractal structure on using CVT of different integer values]

[Figure 4: Shows the fractal generated by the CVT]

[Figure 5: Shows the hierarchy to generate the above fractal]

We have analyzed the table I and found some interesting patterns as follows:
1. Diagonal values are having one type of pattern contains all possible even integers 0, 2, 4, 6, 8…etc.
2. Starting from (0,0) position one can construct a (2x2) matrix by filling the values in its right, bottom and diagonal (bottom-right) positions and recursively this will leads to (4x4), (8x8), (16x16) matrices…etc that can be seen in "Fig. 4".
3. A recursive pattern exists in each square block of size ($2^k$x$2^k$) for k=0, 1, 2, 3…etc. starting from (0, 0) position and if this block is partitioned in the middle (both in row and column) into 4 equal sub blocks of size ($2^{k-1}$x$2^{k-1}$) then also the same pattern can be observed. If the 4-blocks are treated as 4 quadrants (1$^{st}$, 2$^{nd}$, 3$^{rd}$ and 4$^{th}$) in the two dimensional Euclidian space then the patterns in first three quadrants are exactly same and in the 4$^{th}$ quadrant the pattern is same only values are different and that can also be easily constructed by adding $2^k$ to each element of any of the three quadrants. That is if $a_{ij}$ denote the entry of $i^{th}$ row and $j^{th}$ column of the CV-table of size ($2^k$x$2^k$), then

$$a_{ij} = \begin{cases} a_{(i \bmod 2^{k-1})(j \bmod 2^{k-1})} & \text{for } 0 \leq i, j < 2^{k-1} \\ a_{(i \bmod 2^{k-1})(j \bmod 2^{k-1})} + 2^k & \text{for } 2^{k-1} \leq i, j < 2^k \end{cases}$$

4. One can partition a square block of size ($2^k$x$2^k$) for k=0, 1, 2, 3…etc. into two different classes. Where the elements of 1$^{st}$, 2$^{nd}$ and 3$^{rd}$ are in Class-I and the elements of 4$^{th}$ quadrant is in Class-II. Further Class-I elements can be obtained from Class-II by element wise modulo $2^k$ operations ($b = (a_{ij}) \bmod 2^k$). Thus $2^k$ can act as a pivot.
5. Take, a sequence of non-negative integers from $2^n Z$ (n≥0), find the CV table. Observe that for any n≥0, after infinitely many steps we will be having the same fractal what we have got earlier as shown above.

6. Take, an odd sequence i.e. {1, 3, 5, 7, 9, 11, 13, 15…} and try to find the CV table. Observe the pattern of zeros. This pattern is same as what we have found earlier in above. Now, if we multiply the sequence by $2^n$ ($n \geq 0$), then this sequence also lead to the same pattern. Literally, one can say that Sierpinski gasket (fractal) is a very stable fractal.

*Dimension of this fractal*
For this fractal, N=3, S=1/2, where l is the initial length.

Fractal dimension D is given by …      $3 = 1/(1/2)^D$
Or D= log3/log2 ≈ 1.585

This is same as the dimension of Sierpinski triangle. Thus CVT fractal as obtained by us can be regarded as a relative to *Sierpinski* triangle [5].
In [12], eleven lists of ways are given for the construction of *Sierpinski gasket* (also, the Sierpinski triangle) and the author conjectured that there are undoubtedly more. Thus CVT defined in this paper is another way for the construction of this kind of fractal. Next section discusses other various ways to construct the self-similar CVT fractal as obtained in "Fig 3".

## 5. Various ways to construct the CVT fractal

5.1 *L- System to generate CVT fractal*
Let $F_1$, $F_2$ denote horizontal and vertical line segments respectively. Starting from the axiom as shown in table II and "Fig. 5" two rules are defined which produces the above fractal generated by CVT, after infinitely many iterations.

**Table 1.** A table with fractal dimension of fractals according as base of the number system

| L-System for CVT fractal | |
|---|---|
| Variables: | $F_1$, $F_2$ |
| Constants: | - |
| Axiom: | $F_2 - F_1$ |
| Rules: | $F_1 = F_1 - F_2 - - F_2 - F_1$ <br> $F_2 = F_2 - F_1 - - F_1 - F_2$ |
| Angle increment: | 90 degrees |

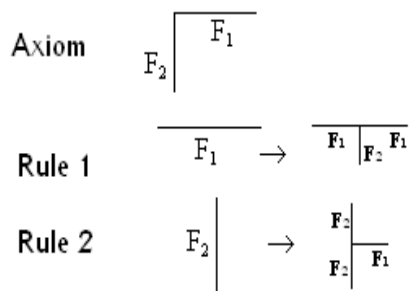

[Figure 6: Ssyntactic representation of Axiom and Rules]

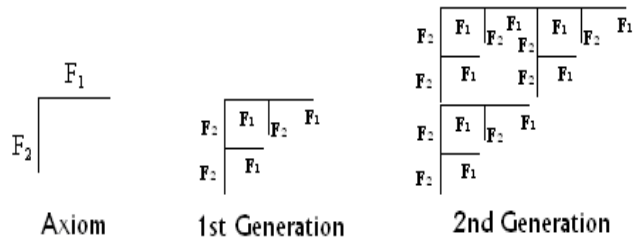

[Figure 7: L-System is used to generate the above fractal]

5.1 *Cellular Automata rules to construct CVT fractal*

In [4] we found that the fractal generated from one-dimensional, two-neighborhood, binary Cellular Automata rule starting from an initial sheet, if rotated by an angle of 180 degree is same as our CVT fractal. According to Wolfram naming convention [10], [11] this is rule number 6, whose truth table and initial sheet on which rule 6 would be applied are given below…

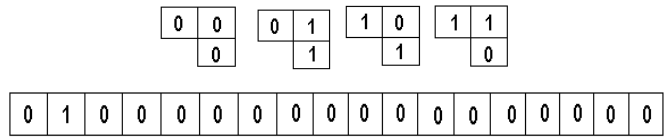

[Figure 8: The rules and the initial seed of CA to generate the CVT fractal]

After finitely many steps, the space-time pattern for the above CA rule is shown in "Fig. 8".

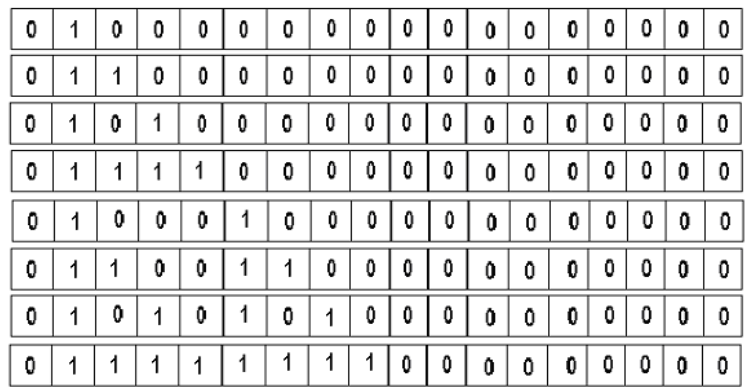

[Figure 9: CA evolution from the initial seed generates the CVT fractal]

"Fig. 9" shows the same CA evolution by replacing 1 by a black cell and 0 by a white cell as usual done by Wolfram. Now by flipping vertically (same as the vertical rotation) this CA evolution we get the same CVT fractal as shown in "Fig. 3".

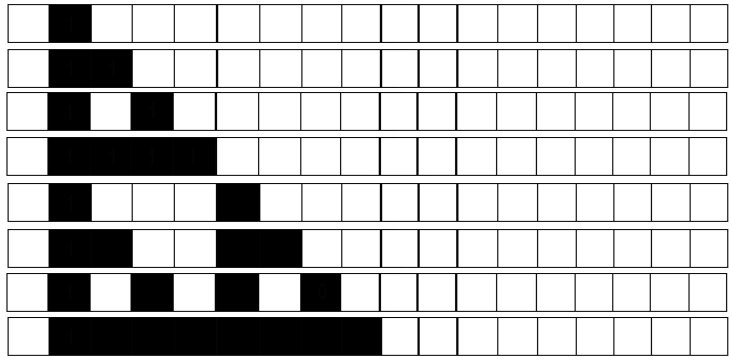

[Figure 10: The rules and the initial seed of CA to generate the CVT fractal]

5.1 *Synthesis of CVT fractal by Tilling*
The constituent parts on synthesis can give rise to the fractal picture. Here we have used four key tiles those are used to generate CVT fractal shown in "Fig. 10".

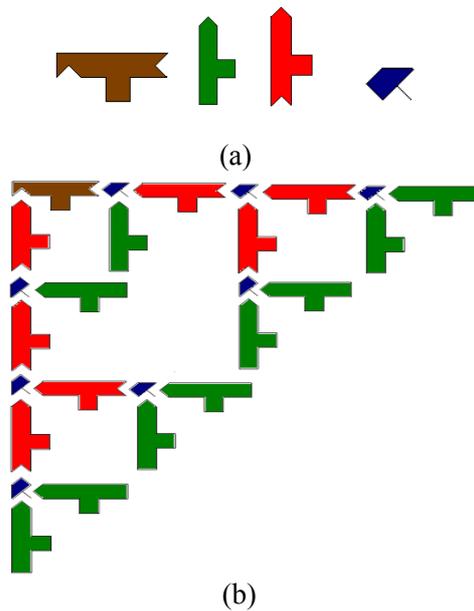

(a)

(b)

[Figure 11: (a) shows four key tiles those are used to generate the CVT fractal and (b) shows the arrangements of these keys.]

Different ways of construction of the self-similar CVT fractal signifies that like other tools CVT can also be used as a construction tool for the formation of different kinds of self-similar fractals. Next section shows that not only the CV-table produces self-similar fractals but CVT can also be used to generate periodic as well as chaotic patterns.

## 5. CVT for generating Periodic and Chaotic patterns

Already we have shown in section 4, that CVT can produce a self-similar pattern (fractal). Here we would like to produce another type of source to have a periodic pattern. For that, first of all we need a periodic sequence of numbers where we can be able to apply CVT to obtain the pattern. And we have a renowned domain of such sequences those are obtained considering decimal representation of rational numbers. Let us consider an example of rational number 1/7. The decimal representation of 1/7 is 0.<u>142857</u>142857142857142857 142857 ... (142857 is being repeated).
Taking the above decimal representation as a sequence of integers, let us construct the CV-Table in binary number system as follows…

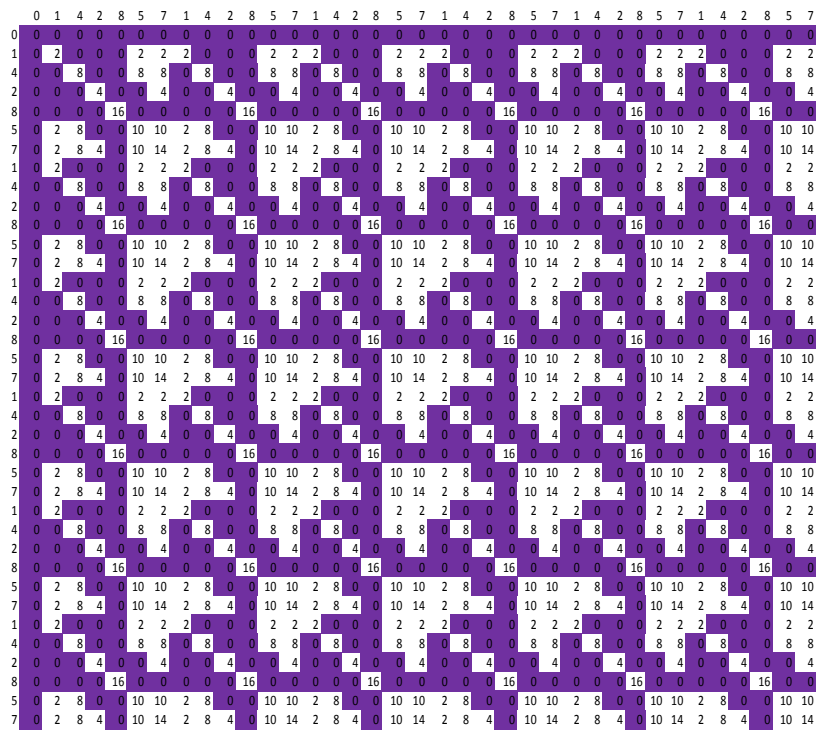

[Figure 12: a fractal structure on using CVT of different integer values]

Here we have considered the pattern of zeros, which we have made it colored. Clearly this pattern of zeros is periodic. Here, we can conclude that for every rational number we can be able to have periodic pattern using CVT, sometimes it may be followed by some chaotic patterns as if starts from an explosion then things becomes stable. That is, here we can encounter a source of countably infinite number of almost periodic patterns corresponding to rational numbers because we know that the set of rational numbers is countable and infinite.

5.1 *Formation of Chaotic Pattern*
We have not yet seen that whether CVT can produce a non-periodic or chaotic pattern or not! Now we are ready to demonstrate an example where we could have a chaotic pattern using CVT.
First of all we need a random sequence of integers, and then we will be applying CVT as we have applied earlier. Here also we have a good domain of such sequences, which can be considered the decimal representation of irrational numbers.
Let us consider an example of an irrational number √2. The decimal representation of √2 is 1.41421356237309504880168872422969807856867187537694807317667973799...(Non-periodic, non-recurring). Considering the above sequence let us try to construct the CV-table in binary number system as shown in table IV.

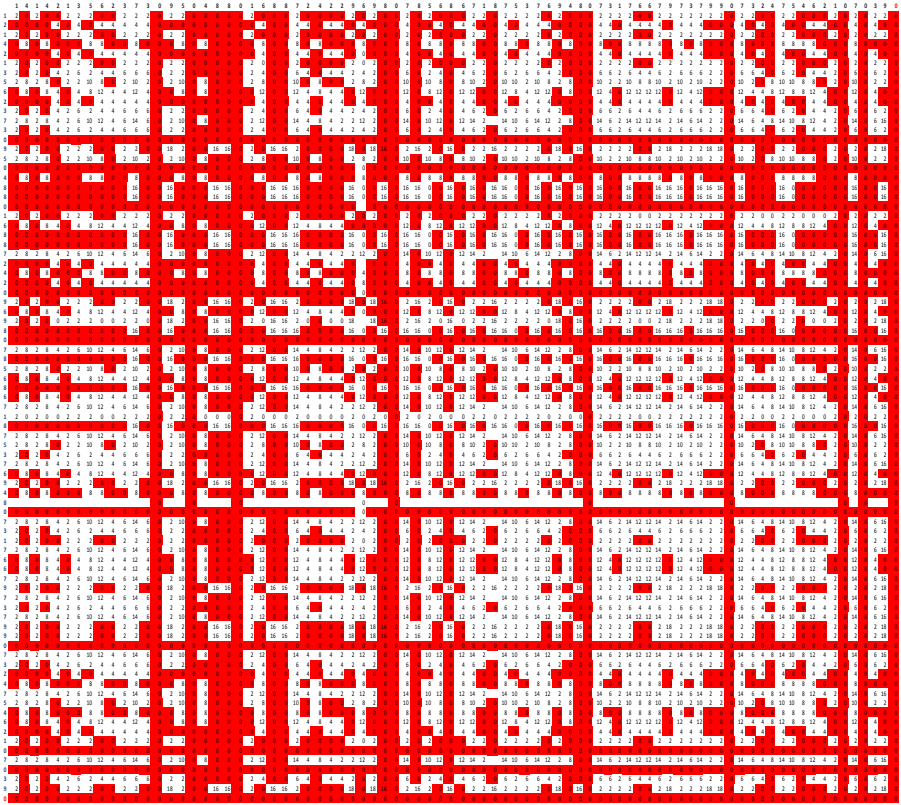

[Figure 13: a fractal structure on using CVT of different integer values]

Here we have considered the pattern of zeros, which we have made it colored. Clearly this pattern of zeros is a non-periodic (chaotic).

Here, we can conclude that for every irrational number we can be able to have non-periodic (chaotic) pattern using CVT. That is here from we can be able to conclude that an uncountable number of periodic and non periodic pattern corresponding to any real numbers can be obtained through CVT. As we know that the set of real numbers is uncountable.

## 5. Analytical and Algebraic properties of CVT

It is shown in section V that, the fractal obtained by CVT can also be constructed by 1-D CA rules, which is basically a sub class of Discrete Dynamical System. This motivates us to study about the dynamical properties of CVT. Interestingly (Z, CVT) is a Discrete Dynamical System. Let us make you recalled the definition of a Dynamical system.

**Dynamical system**: A dynamical system is a semi-group G acting on a space M, i.e. there is a map
$$T: G \times M \to M$$
$$(g, x) \to T_g(x)$$
$$\text{such that } T_g * T_h = T_{g*h} \dots \dots \dots \dots \dots \dots \dots (1)$$

If $(G,*)$ is a group, then $(M, T)$ is called Invertible dynamical system.

In another form we can formalize the condition – (1) as follows.
A dynamical system is semi-group G acting on a space M, i.e. there is a map
$$T: G \times M \to M$$
$$(g, x) \to T_g(x)$$
$$\text{such that there exist a function } f \text{ for which}$$
$$f(T_g, T_h) = T_{f(g,h)} \dots \dots \dots \dots \dots \dots \dots (2)$$

Let us come to in our arena….
Let us define $CVT: Z \times Z \to Z$; here $G = Z, M = Z$
Such that $CVT_a(x) = CVT(a, x)$ $a \epsilon Z, x \epsilon Z$

*Theorem 1*: (Z, CVT) is a Discrete Dynamical System.

Proof: To show (Z, CVT) is a Dynamical system we have to find a function $f$ such that
$f(CVT_a, CVT_b) = CVT_{f(a,b)}$
Let us consider the map $f$ be the CVT. Clearly, $(Z, CVT)$ is a semi-group.
Let a, b, x $\in$ Z and a=$(a_n, a_{n-1}, a_{n-2}, \ldots, a_1)_2$, b=$(b_n, b_{n-1}, b_{n-2}, \ldots, b_1)_2$, x=$(x_n, x_{n-1}, x_{n-2}, \ldots, x_1)_2$"
Now,
$CVT(CVT_a(x), CVT_b(x))$
= $(a_n \wedge x_n, a_{n-1} \wedge x_{n-1}, \ldots, a_1 \wedge x_1, 0) \wedge (b_n \wedge x_n, b_{n-1} \wedge x_{n-1}, \ldots, b_1 \wedge x_1, 0)$
= $((a_n \wedge x_n) \wedge (b_n \wedge x_n), (a_{n-1} \wedge x_{n-1}) \wedge (b_{n-1} \wedge x_{n-1}), \ldots, (a_1 \wedge x_1) \wedge (b_1 \wedge x_1), 0), 0)$
= $((a_n \wedge b_n \wedge x_n), (a_{n-1} \wedge b_{n-1} \wedge x_{n-1}), \ldots, (a_1 \wedge b_1 \wedge x_1), 0)$   i.e.
= $CVT((a_n \wedge b_n, a_{n-1} \wedge b_{n-1}, \ldots, a_1 \wedge b_1, 0), (x_n, x_{n-1}, \ldots, x_1))$
= $CVT(CVT(a,b), x)$
= $CVT_{CVT(a,b)}(x)$
$CVT(CVT_a, CVT_b) = CVT_{CVT(a,b)}$
Hence, (Z, CVT) is a Discrete Dynamical System.
It can be noted that this definition of CVT when treated as a binary operation doesn't satisfy algebraic properties such as Closure, Associative, Existence of identity and Existence of inverse except Commutativity. But as we have seen in section IV CVT could produce fractals. Now we can slightly modify our CVT definition and redefine it so that without affecting the fractal formation it also satisfies some of those interesting algebraic properties.

## 6. Modified Carry Value Transformation (MCVT)

*Formal definition of MCVT*
MCVT is a mapping defined as $MCVT: (B_n \times B_n) \to B_n$ where $B_n$ is the set of strings of length $n$ on $B = \{0,1\}$. More specifically, if $a = (a_n, a_{n-1}, \ldots, a_1)$ and $b = (b_n, b_{n-1}, \ldots, b_1)$ then $MCVT(a,b) = (a_n \wedge b_n, a_{n-1} \wedge b_{n-1}, \ldots, a_1 \wedge b_1)$ is an n bit string and can be computed bit wise by logical AND operation, which denotes no carry, is generated in the LSB at the time of addition procedure.

$$
\begin{array}{rl}
\textit{carry value} = & c_n \quad c_{n-1} \ldots\ldots\ldots\ldots c_1 \\
a = & a_n \quad a_{n-1} \ldots\ldots\ldots\ldots a_1 \\
b = & b_n \quad b_{n-1} \ldots\ldots\ldots\ldots b_1 \\
a \oplus b = & a_n \oplus b_n \quad a_{n-1} \oplus b_{n-1} \ldots\ldots\ldots\ldots a_1 \oplus b_1
\end{array}
$$

[Figure 14: Carry generated in i[th] column is saved in the same i[th] column.]

*Theorem 2*: (MCVT, $B_n$) is a commutative monoid.

**Proof:** As the range of MCVT is $B_n$ so it satisfies *Closure property*. All other properties can be proved as follows.

*For Associative Property*

***Claim:*** $\text{Modified CVT}(\text{Modified CVT}(a,b),c)$
$= \text{Modified CVT}(a, \text{Modified CVT}(b,c))$

*L.H.S*
$\text{Modified CVT}(\text{Modified CVT}(a,b),c)$
$= \text{ModifiedCVT}((a_n \wedge b_n, a_{n-1} \wedge b_{n-1},...,a_1 \wedge b_1),(c_n,c_{n-1},...,c_1))$
$= (a_n \wedge b_n \wedge c_n, a_{n-1} \wedge b_{n-1} \wedge c_{n-1},...,a_1 \wedge b_1 \wedge c_1)$

*R.H.S*
$\text{Modified CVT}(a, \text{Modified CVT}(b,c))$
$= \text{ModifiedCVT}((a_n, a_{n-1},...,a_1),(b_n \wedge c_n, b_{n-1} \wedge c_{n-1},...,b_1 \wedge c_1))$
$= (a_n \wedge b_n \wedge c_n, a_{n-1} \wedge b_{n-1} \wedge c_{n-1},...,a_1 \wedge b_1 \wedge c_1)$

*For Existence of identity*
$\text{ModifiedCVT}((a_n, a_{n-1},...,a_1),(1,1,...,1))$
$= (a_n \wedge 1, a_{n-1} \wedge 1,...,a_1 \wedge 1) = (a_n, a_{n-1},...,a_1)$

*For Commutative*
$\text{Modified CVT}(a,b) = (a_n \wedge b_n, a_{n-1} \wedge b_{n-1},...,a_1 \wedge b_1)$
$= (b_n \wedge a_n, b_{n-1} \wedge a_{n-1},...,b_1 \wedge a_1)$
$= \text{ModifiedCVT}(b,a)$

Hence, (MCVT, $B_n$) is a ***commutative monoid***.

In the modified CV table the decimal values in each entries are exactly half i.e. $a_{ij} = a_{ij}/2$ because only the LSB position 0 is not padded in this type of vectors. Thus the fractals generated by both these CVT operations are exactly same but the modified CVT has an additional advantage of getting some algebraic properties than the conventional CVT. Following section deals with the extension of both CVT as well as MCVT in higher dimensions.

## 7. Extension of CVT and MCVT in higher dimensions

Let us define CVT (or MCVT) recursively in higher dimensional space $Z^k$. CVT (or MCVT) is defined as

$CVT(\text{or } MCVT): (B_n \times B_n \times B_n \times ... \times B_n) \to B_{n+1}$, Where $B_n$ is the set of strings of length $n$ on $B = \{0,1\}$.

More specifically,

$CVT(x_1, x_2,...,x_k) = CVT(CVT(x_1, x_2,...,x_{k-1}), x_k)$

In a similar fashion, we can write

$MCVT(x_1, x_2,...,x_k) = MCVT(MCVT(x_1, x_2,...,x_{k-1}), x_k)$

Where is a positive integer for both CVT and MCVT. It is to be noted that $k = 2$ is the terminating condition for the above recursive procedures.

In particular for $k = 3$, $CVT(x_1, x_2, x_3) = CVT(CVT(x_1, x_2), x_3)$ where $CVT(x_1, x_2)$ could be evaluated as defined above in section 3.

This definition helps us to generate fractals in space, where as earlier we have got fractals in plane. On exploring these ideas we can build up fractals in n-dimensional space.

## 8. Conclusion and Future Research directions

This paper presents a new transformation named as Carry Value Transformation (CVT) applied on a pair of integers. Previously we have used this CVT for Efficient Hardware design of arithmetic operations [2]. On further investigation of this transform in binary number system produces a beautiful pattern, which is found to be a fractal having dimension 1.585, same as that of Sierpinski triangle. Further CVT can be applied for the production of periodic and chaotic patterns. Interestingly, it is proved that **(**Z, CVT) is a Discrete Dynamical System. Further, the definition of CVT is slightly modified and its mathematical properties are highlighted where we have shown that (MCVT, $B_n$ ) is a commutative monoid. Finally, the extension of CVT and modified CVT (MCVT) are done in higher dimensions.

Authors are of firm conviction that CVT/MCVT can be used in the way the other mathematical transforms (e.g., Fourier, Discrete Cosine, Laplace, Wavelet, Cellular Automata Transforms etc.) are used; only the domains will vary from one transform to another. Further studies of algebraic and analytical properties of these transformations are highly needed for the complete exploration of this potential area. Further, authors are expecting to get a close relationship between the CVT/MCVT with many different application areas like Computational geometry, Data compression, Quad trees, Defective chessboards, Pattern Classification, Theory of Computation, and Analysis of Cellular Automata Rules etc. Exploring all these will be our immediate future research directions.

**References:**

[1] B.B. Mandelbrot, *The fractal geometry of nature*. New York,
[2] P.P. Choudhury, S. Sahoo, M. Chakraborty, 2008 Implementation of Basic Arithmetic Operations Using Cellular Automaton, ICIT-08, *IEEE CS Press*, 11[th] International Conference on Information Technology, pp 79-80,
[3] J. L. Hennessy and D. A Patterson, 1996"Computer Architecture: A Quantitative Approach" (**2[nd] edition**), Morgan Kaufmann, San Francisco.
[4] M. Barnsley, 1988 *Fractals Everywhere*, Academic Press Inc. (London) Ltd., San Diego.
[5] H. O. Pietgen, H. Jurgeens, D. Saupe, 1992 *Chaos and Fractals New Frontiers of Science*, ISBN 3-540-97903-4, Springer Verlag.
[6] P. Prusinkiewicz, A. Lindenmayer, 1996 *Algorithmic Beauty of Plants*, ISBN: 0387946764.
[7] S. Wolfram, 1994, *Cellular Automata and Complexity Collected Papers*, Addison-Wesley Publishing Company, ISBN 0-201-62664-0.
[8] S. Wolfram, Statistical Mechanics of Cellular Automata, Reviews of Modern Physics, vol. 55, pp. 601-644, 1983.
[9] C. Bandt, P. Gummelt, Fractal Penrose Tilings Aequationes Math. 53, pp 295-307, 1997.
[10] A. K. Ghosh, P. P. Choudhury, A. Basuray, Chaotic Fractals with Multivalued Logic in Cellular Automata, CISSE 2006, IEEE, University of Bridgeport, Innovations and Advanced Techniques in Computer and Information Sciences and Engineering, pp.77-82, 2007.
[11] A.K. Ghosh, P. P. Choudhury, R. Choudhury, Production of fractals by various means and measuring their dimensions with probable explanation, Laser Horizon, *Journal of Laser Science and Technology Centre* (LASTEC), vol. 6/No. 2, pp.50-59, 2003.
[12] A. Bogomolny, Cut The Knot! An interactive column using Java applets http://www.cut-the-knot.org/ctk/Sierpinski.shtml
[13] P. P. Choudhury, S. Sahoo, B. K Nayak, and Sk. S. Hassan, 2009, Carry Value Transformation: It's Application in Fractal Formation 2009 IEEE International Advanced Computing Conference (IACC 2009), Patiala, India, 6-7 March, pp 2613-2618, 2009.